\documentclass[a4paper,12pt]{article}
\usepackage[utf8x]{inputenc}
\usepackage{graphicx}
\usepackage{amssymb,amsmath}
\usepackage{textcomp}
\usepackage{cite}

\title{Thermodynamics with fractal structure, Tsallis statistics and hadrons}
\author{A. Deppman}
\date{Instituto de Física -  Universidade de São Paulo \\ email: deppman@if.usp.br \\ Rua do Matão Travessa R Nr.187 CEP 05508-090 Cidade Universitária, São Paulo - Brasil}
\begin{document}

\maketitle

\begin{abstract}
 A system presenting fractal structure in its thermodynamical functions is introduced, and it is shown that Tsallis statistics is the correct framework for describing the thermodynamical aspects of such fractal. Its Haussdorf dimension and its Lipshitz-H\"older exponent are determined in terms of the entropic index $q$. The connections with the intermittency in experimental data is discussed. The thermodynamical aspects of the thermofractal is related to the microscopic interaction of its components through the S-matrix.
\end{abstract}

\section{Introduction}

In this work it is shown that a particular thermodynamical system presenting a hierarchy of subsystems, each of them being described by thermodynamical distributions similar or affine to those for the subsystems at different levels of the hierachical structure, is described by Tsallis statistics ($S_q$). Moreover the thermodynamical potential for this system allows a direct connection with the S-matrix for the interacting particles in a gas of thermofractals.

The generalization of Boltzmann statistics proposed by C. Tsallis~\cite{Tsallis1988} has found application in a large number of phenomena in many different fields of knowledge. It is interesting to notice that the main motivation to the introduction of a non additive entropy, $S_q(p)$, which would lead to a non extensive statistics was its applicability to fractal or multifractal systems since this entropy would naturally lead to power-law distributions characteristic of fractals.

The term fractal was coined by Mandelbrot~\cite{Mandelbrot} to designate systems presenting scaling symmetry. For such systems their dimension, according to the definition by Haussdorf, is not necessarily an integer~\cite{Tel}. The definition of fractal can be applied to distribution functions, where the concept of affinity appears. In such cases there are usually many different dimensions associated to the scale symmetry~\cite{Tel, Schroeder} and the system is called multifractal.

There is a large number of fractals found in mathematical relations or in physical systems. Indeed fractals are rather ubiquous, and one reason for such ubiquity may be the fact that complex structures can arise from very simple relations iterated several times. Physics laws are in general simple, so it may be the case that most of the complexity observed in nature emerges from self-similar structures, as it happens with fractals. Quoting Mandelbrot, ``Fatou's and Julia's discoveries confirm in effect, that a very complex artifact can be made with a very simple tool (think of it as a sculptor's chisel), as long as the tool can be applied repeatedly.''~\cite{Mandelbrot}.

The connections between $S_q$ and fractals have already been addressed in other works~\cite{LyraTsallis, Jizba, Touchette, Olemskoi}. In particular, it was argued in Ref.~\cite{Olemskoi} that the statistical mechanics of self-similar complex systems with fractal phase space is governed by Tsallis statistics.

Of special interest for the present work are the thermodynamical aspects of high energy collisions. Such thermodynamical aspects were first observed by Fermi~\cite{Fermi} and subsequently developed by Hagedorn~\cite{Hagedorn65} fifty years ago by supposing a self-similar structure for the hadrons. This was done by the following definition of fireballs
\bigskip

{\it ``fireball is a * statistically equilibrated system composed by an undetermined number of fireballs, each one of them being, in its turn, a (goto *)''}.

\bigskip
 This definition lets explicit the self-similarity of the fireball structure, resulting in a scale invariance typical of fractals, as already mentioned in Ref.~\cite{Beck, WWselfsymmetry}. From the above definition and using a self-consistent argument Hagedorn obtained the complete thermodynamical description of fireballs. Among the predictions were the limiting temperature and the mass spectrum formula, which allowed comparison with experimental data.

Such recursive aspect of the definition was also used by S. Frautschi~\cite{Frautschi} who proposed that {\it hadrons are made of hadrons}. With this definition he was able to derive some of the results obtained previously by Hagedorn. 

Hagedorn's thermodynamical approach was proposed some years before the quark structure of hadrons became accepted, but it had far-reaching consequences. In fact the very idea of a phase transition between the confined-deconfined regimes of hadronic matter was advanced by Cabibbo and Parisi~\cite{CabibboParisi} as a reinterpretation of the limiting temperature discovered with the self-consistent thermodynamics. However, with experiments at higher energies ($\sqrt{s} > 10$~GeV) it was soon noticed that Hagedorn's thermodynamics was not able to describe the transverse momentum ($p_T$) distributions obtained in the High Energy Physics (HEP) experiments. Hagedorn himself proposed a phenomenological model~\cite{Hagedorn83} which gives a power-law distribution that fits data even in the high $p_T$ range, a result impossible to be obtained with his former theory.

In 2000 it was shown that simply changing the exponential function in the self-consistent thermodynamics distribution by the q-exponential function from Tsallis statistics would result in a power-law distribution for $p_T$ which can describe the data in the whole $p_T$ range~\cite{Beck, Bediaga2000156}. In 2012 the self-consistent principle proposed by Hagedorn was generalized by the inclusion of $S_q$ leading to a well-defined thermodynamical theory when Boltzmann statistics is replaced by Tsallis statistics~\cite{Deppman}. In this case not only the temperature $T$ must be constant, but also the entropic index $q$ from the non extensive statistics must be independent of the collision energy or of the hadron mass. In addition a new formula for the hadron mass spectrum is obtained in terms of the q-exponential function, where the parameters $T$ and $q$ can be determined.

In the last few years several experimental data from HEP have been analysed using the thermodynamical formula derived from Tsallis statistics~\cite{Sena,Lucas,Cleyman_Worku, AzmiCleymans1, AzmiCleymans2} or using the power-law formula inspired on QCD~\cite{Alice2012, CMS2011,WongWilk, Wilk2007, Bonasera}. In a recent work~\cite{Lucas} it was shown that both formulas fit the $p_T$-distribution data very well, but the parameters obtained from the fitting procedure present very different behaviours with energy or particle mass. When the thermodynamical formula is used both $T$ and $q$ are independent of the collision energy and on the mass of the particle analysed. In addition, it was shown that $T$ and $q$ obtained from the analysis of the mass spectrum are consistent with those obtained with the analyses of $p_T$-distribution. In this context it is remarkable that the new mass formula proposed in Ref.~\cite{Deppman} fits well even the region of mesons as light as pions.

The subject remains controversial. From one side there is the idea that a thermodynamical approach based on the non extensive statistics can describe the data in the whole $p_T$ range with parameters $T$ and $q$ which are not only independent of energy and mass, as demanded by the non extensive self-consistent thermodynamics\cite{Deppman}, but also present values that are in accordance with a completely different analysis based on the mass spectrum of hadrons~\cite{Lucas}. On the other hand the power-law appoach inspired in QCD presents the advantage of being more closely related to the fundamental interactions of hadrons~\cite{Wong}.

In this context it is important to investigate the possible origins of non extensivity in QCD. There are some connections between Tsallis and Boltzmann statistics already proposed, as
  \begin{itemize}
   \item Particular case of Fokker-Plack equation~\cite{Borland}.
   \item Temperature fluctuation in a stationary state~\cite{Beck2,Wilk2009}.
   \item Finite size of thermodynamical systems~\cite{Biro}.
  \end{itemize}
These approaches triggered an interesting discussion around more general definitions of entropy, like in superstatistics~\cite{BeckCohen} or in formulations of new entropies based on relaxation of the four Shannon-Khinchin axioms~\cite{Thurner,Tempesta}. The connections between Boltzmann and Tsallis statistics proposed so far, however, are related to thermodynamical aspects of the system but are not directly related to the microscopic aspects of hadronic matter and to QCD interaction.

A comparison of results from the non extensive self-consistent thermodynamics and from Lattice QCD (LQCD) has been performed~\cite{Deppman2} showing a fair agreement between the two methods. Since LQCD calculations do not include explicitly the non extensive features present in the thermodynamical calculations, one can understand from here that non extensivity must be an emergent characteristic from the QCD interaction in systems like those obtained in HEP experiments. A recent work~\cite{Wong} used a phenomenological model based on first-order calculation of the parton-parton cross section to obtain a power-law behaviour describing the $p_T$-distributions even at low values of transverse momentum, what was attributed by the authors to a dominance of hard-scattering. These are indications that one could learn about QCD from the non extensive features of the experimental distributions.

The present work addresses the possibility of finding close relations between the non extensive thermodynamics and the fundamental QCD interaction of hadrons. To this end a system showing a fractal structure in its thermodynamics will be introduced and its relation with Tsallis statistics will be deduced. Some features of this system will be studied and finally a relation between the entropic index, $q$, and the $S$-matrix for the interacting gas of quantum system will be obtained.

This paper is organized as follows: in section 2 some well-known results for an ideal gas are reviewed, and them they are used in section 3, where a system described by a fractal-like thermodynamics is defined where the constituent parts of this system have an internal structure which is similar to that of the main system, like the fireballs defined by Hagedorn. Then it is shown that this system present self-affine distributions that characterize multifractals. In addition it is possible to obtain a system with self-similar distributions. It is shown that in both cases the Tsallis statistics is the most natural statistics to describe the thermodynamical aspects of such systems. In section 4 the main features of that system are discussed, as its fractal chraracteristics. The results are then used to investigate the possibility of fractal tructure in hadrons, when experimental data on intermittency in multiparticle production in HEP is used to corroborate the hypothesis used here. Finally the connection between S-matrix and non extensivity is established. In section 5 the conclusions of this work are presented.

\section{Energy fluctuation of an ideal gas}

It is well known that the total energy of an ideal gas fluctuates according to~\cite{Landau}
\begin{equation}
 P(U) dU = A \exp \bigg(- \sum_{i=1}^{3N} \frac{p_i^2}{2mkT} \bigg) d^{3N}p_i 
\end{equation}
where $P(U)\,dU$ is the probability to find the energy of the system between $U$ and $U+dU$, $m$ is the mass of individual particles of the gas and 
\begin{equation}
 d^{3N}p_i = d_{1x} \, d_{1y} \, d_{1z} \dots  d_{Nx} \, d_{Ny} \, d_{Nz}
\end{equation}
is an infinitesimal volume in the momentum space. $A$ is a normalization constant which can be straightforwardly determined, giving
\begin{equation}
 A=(2 \pi m k T)^{-\frac{3N}{2}} \,.
\end{equation}

The infinitesimal volume can be written also in terms of the total momentum
\begin{equation}
 p^2=\sum_{i=1}^{3N} p_i^2
\end{equation}
by noticing that
\begin{equation}
 d^{3N}p_i \sim p^{3N-1}dp
\end{equation}
where $p$ is the radius of a hypersphere in a $3N$-dimensinal space. Of course
\begin{equation}
 U=\frac{p^2}{2m}\,,
\end{equation}
therefore
\begin{equation}
 \frac{dU}{U}=\frac{dp}{p}\,.
\end{equation}
From Eqs. 1-7 it is possble to conclude that
\begin{equation}
 P(U) dU= (kT)^{-\frac{3N}{2}} U^{\frac{3N}{2}-1} \exp\bigg(-\frac{U}{kT}\bigg) dU\,, \label{Maxwell}
\end{equation}
which is consistent with the Maxwell distribution of velocities. Note that Eq.~(\ref{Maxwell}) does not depend explicitly on the particle mass.

Based on this result for the ideal gas the thermofractal system will be introduced in the next section.

\section{Thermodynamics with fractal structure}

Define thermofractal as a class of thermodynamical systems presenting a fractal structure in its thermodynamical description in the following sense:
\begin{enumerate}
 \item The total energy is given by
 \begin{equation}
  U=F+E\,,
 \end{equation}
where $F$ corresponds to the kinetic energy of $N'$ constituent subsystems and $E$ corresponds to the internal energy of those subsystems, which behaves as particles with an internal struture.

\item The constituent particles are thermofractals. The ratio $\langle E \rangle /\langle F \rangle$ is constant for all the subsystems. However the ratio $E/F$ can vary according to a distribution, $\tilde{P}(E)$, which is self-similar (self-affine), that is, at different levels of the subsystem hierarchy the distribution of the internal energy are equal (proportional) to those in the other levels. 

\item At some level $n$ in the hierarchy of subsystems the phase space is so narrow that one can consider 
\begin{equation}
 \tilde{P}(E_n) d\,E_n=\rho dE_n\,,
\end{equation}
with $\rho$ being independent of the energy $E_n$.
\end{enumerate}

For the description of the thermodynamical properties of such a system the starting point is the Boltzmann factor
\begin{equation}
 P(S)=A\,\exp(-S/k)\label{entropydistr}
\end{equation}
with $S$ being the entropy and $k$ the Boltzmann constant. Supposing the variations of the volume can be disregarded one has
\begin{equation}
 dU=T\,dS\,,
\end{equation}
so the probability in Eq.~(\ref{entropydistr}) can be written in terms of the total energy as
\begin{equation}
 P(U) dU=A \exp(-U/kT) DU\,,
\end{equation}
where $DU$ is a generalized differential. Due to properties 1 and 2 of thermofractals one  has
\begin{equation}
 P(U) dU=A \exp(-\alpha F/kT) DF DE \label{FEprob}
\end{equation}
with 
\begin{equation}
 \alpha=1+\frac{\varepsilon}{kT} \label{alfa}
\end{equation}
where %$N$ is an effective number of subsystems which includes internal degrees of freedom, as shown below, and
\begin{equation}
 \varepsilon=\frac{E}{F} kT\,. \label{ratioEF}
\end{equation}
% Notice that $\varepsilon/N$ rescales the effective energy $\varepsilon$ to that of the internal energy of each subsystem.

Since $F$ is related to the kinetic energy part of the constituent particles it is reasonable to write, based on Eq.~(\ref{Maxwell}),
\begin{equation}
 D F= F^{\frac{3N'}{2}-1}  dF
\end{equation}
and for the internal energy it is possible to write
\begin{equation}
 D E= \tilde{P}(E) dE\,,
\end{equation}
where $\tilde{P}(E)$ is the probability density for the subsystem internal energy.

Note that due to Eq.~(\ref{ratioEF}) one has
\begin{equation}
 \tilde{P}(E) dE=\frac{F}{kT} \tilde{P}(\varepsilon)d\varepsilon
\end{equation}
so Eq.~(\ref{FEprob}) is now given by
\begin{equation}
 P(U) dU= A F^{\frac{3N}{2}-1} \exp\bigg(-\frac{\alpha F}{kT}\bigg) dF \tilde{P}(\varepsilon) d\varepsilon\,,
\end{equation}
where $N=N'+2/3$ is an effective number of subsystems. Factors not depending on $\varepsilon$ or $F$ are included in the constant $A$.

The thermodynamical potential is given by
\begin{equation}
 \Omega=\int_0^{\infty} A F^{\frac{3N}{2}-1} \exp\bigg(-\frac{\alpha F}{kT}\bigg) dF \tilde{P}(\varepsilon) d\varepsilon\,. \label{Omega}
\end{equation}
which, after integration on $F$ results in
\begin{equation}
 \Omega= \int_0^{\infty} A \bigg[1+\frac{\varepsilon}{kT}\bigg]^{3N/2} \tilde{P}(\varepsilon) d\varepsilon \,. \label{Oeps}
\end{equation}

\subsection{Self-affine solution}

Now using property 2 it can be imposed the self-affinite in the probability functions by establishing
\begin{equation}
 \log[P(U)] \propto \log[\tilde{P}(\varepsilon)]\,. \label{selfaffinite}
\end{equation}
Eqs.(~\ref{Oeps}) and~(\ref{selfaffinite}) are simultaneously satisfied if
\begin{equation}
 P(\varepsilon)= A \bigg[1+\frac{\varepsilon}{kT}\bigg]^{-\frac{3Nn}{2}}\, \label{selfaffine}
\end{equation}
where $n$ is the number of levels in the subsystem hierarchy according to property 3.

Defining
\begin{equation}
 q_n -1=\frac{2}{3Nn}
\end{equation}
and 
\begin{equation}
 \tau=(q_n-1)T
\end{equation}
it is finally obtained
\begin{equation}
 P_n(\varepsilon)=A\bigg[1+(q_n-1)\frac{\varepsilon}{k\tau} \bigg]^{-\frac{1}{q_n-1}}\,.
\end{equation}
which is the well-known Tsallis distribution. Notice that this system presents several entropic indexes $q_n$ depending on the hierarchical level $n$ of the thermofractal. In the next section it will be shown that it is possible to obtain a thermofractal with  $q$ independent of the fractal level.

\subsection{Self-similarity}

By slightly modifying Eq.~(\ref{Oeps}) and writing
\begin{equation}
 \Omega= \int_0^{\infty} A \bigg[1+\frac{\varepsilon}{kT}\bigg]^{-\frac{3N}{2}} [P(\varepsilon)]^{\nu} d\varepsilon\,, \label{Omegasimilar}
\end{equation}
where $\nu$ is a fractal index, it is possible to impose the identity
\begin{equation}
 P(U) = \tilde{P}(\varepsilon)\, \label{similarity}
\end{equation}
corresponding to a self-similar solution for the thermofractal probability distributions. The simultaneous solution for Eqs.~(\ref{Omegasimilar}) and~(\ref{similarity}) is obtained with
\begin{equation}
 P(\varepsilon)= A \bigg[1+\frac{\varepsilon}{kT}\bigg]^{-\frac{3N}{2}\frac{1}{1-\nu}}\, \label{selfsimilar}\,.
\end{equation}

Introducing the index $q$ by
\begin{equation}
 q-1=\frac{2}{3N} (1-\nu) \label{eq:qnu}
\end{equation}
and the effective temperature
\begin{equation}
 \tau=\frac{2(1-\nu)}{3}T \label{eq:tauT}
\end{equation}
one finally obtains
\begin{equation}
 P(\varepsilon)=A \bigg[1+(q-1)\frac{\varepsilon}{k\tau}\bigg]^{-\frac{1}{q-1}}\,, \label{Tsallis_weight}
\end{equation}
which is exactly the Tsallis q-exponential factor characteristic of the nonextensive statistics.

Eq.~(\ref{Tsallis_weight})  shows that instead of the Boltzmann statistical weight, the Tsallis statistical weight given by the q-exponential function should be used to describe more directly the thermodynamics of thermofractals. In fact, writing
\begin{equation}
 \frac{\langle \varepsilon \rangle}{\tau}=S_q
\end{equation}
it follows from Eq.~(\ref{Tsallis_weight}) that
\begin{equation}
 \frac{S_q}{k}=\frac{1-\sum_i P_i^{1-q'}}{q'-1}\,,
\end{equation}
which is the Tsallis entropy with $q'=2-q$, with $P_i$ representing a discretizated probability based on Eq.~(\ref{Tsallis_weight}). Notice that the change $q \rightarrow q'$ is necessary due to the different definition of the q-exponential used here (see for instance~\cite{Megias}). This result is in agreement with the findings in Ref.~\cite{Olemskoi}, where it is shown that self-similarity in fractal systems are described by Tsallis statistics.

Note that from Eqs.~(\ref{eq:qnu}) and~(\ref{eq:tauT}) one has
\begin{equation}
 q-1=\frac{1}{N} \frac{\tau}{T}\,, \label{eq:q}
\end{equation}
showing that the entropic index $q$ is related to the ratio between the Tsallis temperature $\tau$ and the Hagedorn temperature $T$.

\section{Discussion}

In order to make clear the structure of the thermofractal it will be interesting to analyse what happens when one considers the first level after the initial one in the fractal structure. From Eq.~(\ref{Omegasimilar}) one has
\begin{equation}
 \Omega= \int_0^{\infty} \int_0^{\infty} A F^{\frac{3N'}{2}-1} \exp[-F/kT] \bigg[\exp[-\gamma(\varepsilon)F/kT] [P(\varepsilon)]^{\nu} d(F \varepsilon)\bigg] dF \label{Omegasimilar2}
\end{equation}
where
\begin{equation}
 \gamma(\varepsilon)=\frac{\varepsilon}{kT}\,.
\end{equation}
Considering that $\gamma(\varepsilon)F=E$ and that $d(F \varepsilon)=dE$, one can see that the term between brakets is the internal energy distribution. Considering the the internal energy is distributed statistically among the $N$ constituent subsystems, and considering that they are independent of each other it is possible to write
\begin{equation}
 dE=dE_1 \dots dE_N'
\end{equation}
and
\begin{equation}
 P(\varepsilon)=P_1(\varepsilon) \dots P_N'(\varepsilon)
\end{equation}
with $E_i$ and $P_i$ corresponding to the energy and the probability density for the $i$th subsystem, respectively.

Due to properties 1 and 2 of thermofractals all density distributions are identical, since here the self-consistent solution is under consideration\footnote{For the self-affine solution a similar reasoning can be applied.}, therefore Eq.~\ref{Omegasimilar2} can be written as
\begin{equation}
 \Omega= \int_0^{\infty}  A F^{\frac{3N'}{2}-1} \exp[-F/kT] \bigg\{\int_0^{\infty} \exp\bigg[-\sum_i E_i/kT\bigg] \bigg[\prod_i P_i(\varepsilon)\bigg]^{\nu} dE_1 \dots dE_N'\bigg\} dF\,.
\end{equation}
%where terms dependent on $N'$ are included in the constant $A$.

 The kinetical energy $F$ can be written in terms of the individual subsystems, as describe above in the case of an ideal gas, resulting
 \begin{equation}
  \Omega=\bigg\{\int_0^{\infty} A F_i^{\frac{3}{2}-1} \exp[-F_i/kT] \bigg[\int_0^{\infty} \exp[-E_i/kT] [P_i(\varepsilon)]^{\nu} dE_i\bigg] dF_i\bigg\}^{N'} \,, \label{fracdist1}
 \end{equation}
 with $F_i$ being the kinetical energy of the $i$th subsystem, with $F=\sum_i F_i$.
 
 Notice that the term between square brakets represents the internal energy distribution of one subsystem of the original thermofractal. Therefore, according to property 1, the subsystem is also a thermofractal and due to property 2 its energy $E_i$ can be separated into two parts, $E_i=F'_i+E'_i$, with $F'_i$ being the kinetic energy of the components of the subsystem and $E'_i$ their internal energy. Then
 \begin{equation}
  \int_0^{\infty} \exp[-E_i/kT] [P_i(\varepsilon)]^{\nu} dE_i=\int_0^{\infty} \int_0^{\infty} \exp[-\alpha F'_i/kT] [P_i(\varepsilon)]^{\nu} d\epsilon dF'_i\label{fracdist2}
 \end{equation}

The equation above shows that it is possible to factorize the probability distributions of each subsystem, and it explicitly shows that each of them have an internal energy distribution that has the same form of the original system, according to Eq.~(\ref{Omegasimilar}).

In Eq.~(\ref{Tsallis_weight}), $A$ is a normalizing constant, which gives
\begin{equation}
 A=\frac{2-q}{k\tau}\,.
\end{equation}

The average energy of the thermofractal is then
\begin{equation}
 \langle \varepsilon \rangle = A \int_0^{\infty} \varepsilon \bigg[1+(q-1)\frac{\varepsilon}{k\tau}\bigg]^{-\frac{1}{q-1}}\, d \varepsilon\,,
\end{equation}
resulting
\begin{equation}
 \langle \varepsilon \rangle = \frac{k\tau}{3-2q}\,.
\end{equation}

 From Eq.~(\ref{ratioEF}) and the mean value for $\varepsilon$ it results that
 \begin{equation}
  \frac{\langle \varepsilon \rangle}{kT}=\frac{\langle E \rangle}{\langle F \rangle}=\frac{q-1}{3-2q}N\,. \label{energyratio}
 \end{equation}
 Considering also Eq.~(\ref{eq:q}) it is possible to observe that while the temperature $\tau$ regulates the average energy of the system, the temperature $T$ regulates the ratio between the kinetic energy, $F$, and the internal energy, $E$.
 
  Defining $r=\langle E \rangle/\langle F \rangle$ it is possible to write the ratio
  \begin{equation}
   R=\frac{\langle E \rangle/N'}{\langle U \rangle}=\frac{r/N'}{1+r}\,,
  \end{equation}
 and using Eq.~(\ref{energyratio})
 it is obtained
\begin{equation}
 R=\frac{(q-1)N/N'}{3-2q+(q-1)N}\,, \label{scale}
\end{equation}
which represents the ratio between the internal energy of one of the thermofractal constituent subsystems and the total energy of the main fractal.

It is known that as $q \rightarrow 1$ Tsallis statistics approach Boltzmann statistics, so it is interesting to analyse the thermofractal in that limit. Due to Eq.~(\ref{eq:qnu}), as $q \rightarrow 1$ also $\nu \rightarrow 1$, and from Eq.~(\ref{eq:tauT}) one notice that there are two ways to get this limit: one by letting $\tau \rightarrow 0$ and the other keeping $\tau$ constant.

In the case $\tau \rightarrow 0$ the Boltzmann limit is not obtained. In fact in this case one has $\nu \rightarrow 1$, as in the case of the self-affine solution, but with  $q$ independent of the hierachical level. This is possible only for $\tau \rightarrow 0$ corresponding to the trivial case of a thermofractal with energy $U \rightarrow 0$. This also indicates that the self-similar solution is not a special case of the self-affine solution, but represents a different system.

The Boltzmann limit is obtained if $\tau$ is constant, what means that $(1-\nu)T$ remains constant as $\nu \rightarrow 1$, therefore $T \rightarrow \infty$. Hence the Boltzmann limit is obtained if almost all energy of the gas appears in the form of kinetical energy of its constituents. In this case the system is insensitive to the subsystem internal energy, behaving therefore as an ideal gas that can be described by Boltzmann entropy. 

\subsection{Thermofractal dimensions}

\bigskip
{\bf Haussdorf dimension}

 Consider a hypothetical experiment where the energy of the thermofractal is measured with resolution $r$. This means that energy fluctuations  smaller than $r$ can be neglected, defining in this way the level $n$ of the thermofractal structure where the subsystems internal degrees of freedom can be ignored, according to property 3 above. The level $n$ is such that $R^{n}=r$, so
\begin{equation}
 n=\frac{\log r}{\log R}\,. \label{nratio}
\end{equation}

The Haussdorf fractal dimension $D$~\cite{Tel, Schroeder} is determined by considering that when the energy is measured in units of $r$ the total energy scales as $r^{-1}$ while the energy of each subsystem scales as $r^{-D}$ such that
\begin{equation}
 \mathcal{N}r^{-D}\propto r^{-1}\,,
\end{equation}
where $\mathcal{N}$ is the number of boxes necessary to completely cover all subsystem energies of a thermofractal. It follows the well-known relation
\begin{equation}
 D-1=\frac{\log \mathcal{N}}{\log r}\,.
\end{equation}
Since at the level $n$ all subsystems have distinguishable energies at the given resolution then $N$ is the number of subsystems at this level, i.e., $\mathcal{N}=N'^n$. From here it follows
\begin{equation}
 D=1+\frac{\log N'}{\log R}\,. \label{fracdim0}
\end{equation}

\bigskip
{\bf Fractal spectrum}

There are several parameters that characterize multifractals, and in the following some of those will be investigated. Among these quantities, the Lipshitz-H\"older mass  exponent and the fractal spectrum are the most used~ \cite{Schroeder}. In this context the probability $p(x_i)$ for the event $x_i$ is related to the mass exponent $\alpha_i$ by 
\begin{equation}
 p_{\delta}(x_i) \propto \delta^{\alpha_i}\,,
\end{equation}
where $\delta$ is the linear dimension of the basic box in which the phase space is partitioned.

The partition function
\begin{equation}
 Z(\tilde{q})=\sum_i p_{\delta}^{\tilde{q}}(x_i) \sim \sum_i \delta^{\tilde{q}\alpha_i}\,.
\end{equation}
This partition function is also written in another form
\begin{equation}
 Z(\tilde{q})=\sum_{\alpha_i} \delta^{\tilde{q}\alpha_i} \eta(\alpha_i)\,,
\end{equation}
with 
\begin{equation}
 \eta(\alpha_i) \propto \delta^{-f(\alpha_i)}
\end{equation}
so that
\begin{equation}
 Z_{\delta}(\tilde{q}) \propto \delta^{t(\tilde{q})} \label{LH_def}
\end{equation}
where\footnote{The usual notation is $\tau(q)$ but here it is made use of $t(\tilde{q})$ to avoid confusion with the Tsallis temperature and the entropic index.}
\begin{equation}
 t(\tilde{q})=\tilde{q}\alpha_i-f(\alpha_i)\,, \label{tau}
\end{equation}
using, for the sake of simplicity, $\alpha_i=\alpha(x_i)$. The function $f(\alpha)$ is the multifractal spectrum.

Let us consider the thermofractal which presents a probability density given by Eq.~(\ref{Tsallis_weight}). In order to avoid confusion with the symbols used for probability we will indicate it by $\rho(x)$, with $x=1+\varepsilon/k\tau$. One has
\begin{equation}
 \rho(x) \propto x^{\frac{1}{q-1}} \label{rho}
\end{equation}
so the probability to find particles in the box with dimension $\delta$ around $x$ is
\begin{equation}
 p(x) = \mathcal{N} \rho(x) \Delta x \propto x^{-\frac{1}{q-1}} \delta \sim \delta^{\alpha(x)}
\end{equation}
with
\begin{equation}
 \mathcal{N}=N'^{n}\,,
\end{equation}
It follows that the mass exponent, $\alpha(x)$, is
\begin{equation}
 \alpha(x)-1= n \frac{\log N'}{\log \delta}-\frac{1}{q-1} \frac{\log x}{\log \delta}\,. \label{alfa}
\end{equation}
Using Eq.~(\ref{nratio}) it results
\begin{equation}
 \alpha(x)-1= \frac{\log N'}{\log R}-\frac{1}{q-1} \frac{\log x}{\log \delta}\,. \label{alfa}
\end{equation}

The fractal spectrum is related to the number of boxes with the same index $\alpha$. Therefore consider the probability
\begin{equation}
 \Delta p(x)= \mathcal{N} \rho(x) \delta \Delta x \,. \label{deltap}
\end{equation}
Now the number of boxes with dimension $\delta$ corresponding to the interval $\Delta x$ is given by the relation
\begin{equation}
 \Delta x=\eta(x)  \delta\,.
\end{equation}
Using this result in Eq.~(\ref{deltap}) and considering Eq.~(\ref{rho}) it is obtained
% \begin{equation}
%  \eta(x)=\frac{1}{\delta} \sqrt{\frac{\Delta p}{\rho'(x)}}
% \end{equation}
% and from Eq.~(\ref{rho}) one gets
\begin{equation}
 \eta(x) \delta=  x^{\frac{1}{q-1}} \mathcal{N}^{-1}   \,.
\end{equation}
From the equation above one can see that
\begin{equation}
 \eta(x)=\delta^{-f(\alpha)}
\end{equation}
with
\begin{equation}
 f(\alpha)-1 \log \delta=  n \log N' -\frac{1}{q-1} \log x\,.
\end{equation}
Applying Eq.~(\ref{nratio}) one gets
\begin{equation}
 f(\alpha)-1 = \frac{\log N'}{\log R}-1 -\frac{1}{q-1} \frac{\log x}{\log n \delta}\,. \label{f(alfa)}
\end{equation}
Comparing Eqs.~(\ref{alfa}) and (\ref{f(alfa)}) it results that
\begin{equation}
 f(\alpha)=\alpha\,.
\end{equation}
Note that this result was already expected from the multifractal dimension theory~\cite{Tel, Schroeder}. Also $\alpha$ corresponds to the Haussdorf dimension given in Eq.~(\ref{fracdim0}).

In the limit $\delta \rightarrow 0$ it results
\begin{align}
%\bigg\{
 \begin{cases}
  &f(\alpha)=\alpha=D \\
  &D=1+\frac{\log N'}{\log R} \,. \label{fracspec}
 \end{cases}
\end{align}

The calculations performed here are valid everywhere but fot hte case of $\alpha$ corresponding to the lowest range of probabilities, which is indicated by $\alpha_{max}$. Due to the asymptotic behavior of the probability density one has $p(x\rightarrow \infty)\rightarrow 0$ so $\alpha_{max}\rightarrow \infty$ and also the number of boxes $\eta(x\rightarrow \infty)\rightarrow \infty $, hence $f(\alpha)\rightarrow \infty$. But since the probability does not diverge one has
\begin{equation}
 Z_{\alpha_{max}}=\delta^{\alpha_{max}-f(\alpha_{max})}\rightarrow 0\,,
\end{equation}
therefore $\alpha_{max}-f(\alpha_{max}) \rightarrow \infty$.

The Lipshitz-H\"older exponent is given by Eq.~(\ref{LH_def}). With the results obtained so far one has
\begin{equation}
 Z(\tilde{q})=\delta^{\tilde{q}\alpha-f(\alpha)}
\end{equation}
so
\begin{equation}
 t(\tilde{q})=(\tilde{q}-1)\alpha\,. \label{LHexp}
\end{equation}
The exponent $t(\tilde{q})$ can be observed experimentally, as discussed below.

\subsection{Thermofractals and hadrons}

Before considering to use the thermofractal to get some knowledge about the hadron structure a few comments are needed. In the construction of the thermofractal formalism antisymmetrization was not taken into account. The effects of antisymmetrization however are expected to be small~\cite{Hagedorn65, Frautschi} since the phase space is sufficiently large to consider the hadronic states of interest as a dilute gas.

Another aspect is that the treatment used here is semi relativistic, with the energy of the particles calculated as
\begin{equation}
 E=\frac{p^2}{2m}+m\,
\end{equation}
where the internal energy is identidied with the subsystem mass, $m$. This may be a good approximation when the temperature the temperature $T$ is small so that $E$ is sufficiently larger than $F$.

The formalism derived in the last section is very general even though it has been motivated by the definitions of hadrons given by Hagedorn~\cite{Hagedorn65} and Frautschi~\cite{Frautschi}. Many aspects of this system can be investigated, as its fractal dimensions or its thermodynamical functions. In what follows some aspects of the fractal structure and its phase-space occupation will be addressed, as well as a possible connection between the microscopic interaction of the constituents of the thermofractal and the entropic index $q$ which characterizes its nonextensive statistics. Further analyses on the fractal structure or the possible implications of this formalism on the study of QCD in high energy collisions will be given in future papers. From now on it is supposed that hadrons have a thermofractal structure.

\subsubsection{Hadron fractal dimension}

In order to calculate the fractal properties of hadrons one needs two parameters that characterize the hadronic thermodynamics, namely, the ratio $\tau/T$ and the entropic index $q$. These values have been thoroughly investigated in analyses of $p_T$ distributions from high energy $pp$ collisions~\cite{Cleyman_Worku, Cleymans, Sena, Lucas}, in an analysis of the hadronic mass spectrum~\cite{Lucas}, and in the comparison of the thermodynamical calculations with LQCD data~\cite{Deppman2}. The values found are $q=1.14$ and $\tau/T=0.32$~\cite{Deppman2, Megias}.

Proceeding to calculate the thermofractal properties one has, using Eq.~(\ref{eq:q}), $N=2.3$, and using $N=N' +2/3$ it results $N' =1.7$. From Eqs~(\ref{scale}) one has $R=0.104$. Finally, using Eq.~(\ref{fracdim0}) it results $D=0.69$, so from Eq.~(\ref{fracspec}) also $\alpha=0.69$.

The exponent $t(\tilde{q})$ can be observed experimentally through the intermittency in experimental data, which has been studied in many works on high energy collisions~\cite{Hwa, HwaPan, Bialas_Peschanski, Bialas_Peschanski2,DreminHwa}. Intermittency allows a direct measure of that exponent and has been used as an indication of fractal aspects in multiparticle production. The value calculated here is in fair agreement with the results of analyses of experimental data in hadron-hadron collisions\cite{Gang, Gang2, Gang3, Ajienko, Rasool}, which range between 0.43 and 0.65.

The agreement described above needs to be discussed in more details. In fact, the analysis of intermittency is made through a sophisticated methodology that has been developed some decades ago to extract fractal parameters from experimental data~\cite{Bialas_Peschanski, Bialas_Peschanski2, Hwa, HwaPan} and has being applied since then to study mainly data from heavy ion collisions in emulsion~\cite{Albajar, Ghosh, Ahmad, SinghJain}. But aside the technical difficulties, there is the unavoidable problem described in Ref.~\cite{LipaBuschbeck, Sarkisyan1} where it is shown that when multiple fractal sources are present the measured intermittency is weaker than the real fractal dimension would imply. In fact experimental data where one supposed fewer sources are present tends to present stronger intermittency effects when measured with the available technique. This may explain the fact that the intermittency in nucleus-nucleus collisions, which is $\sim$0.97, are much weaker than that from hadron-hadron or $e⁺e⁻$ collisions, which is $~\sim$0.4~\cite{Gang}.

The fair agreement found between calculation and the experimental values indicates that the thermofractal proposed here can indeed give a reliable description of the fractal aspects of the multiparticle production. In addition, it can show that the intermittency found in HEP data is related to the fractal structure of the hadron. In fact it is the hadron structure of the hadron that leads to the non extensive self-consistent thermodynamics~\cite{Deppman} as the proper thermodynamical description of the hadronic systems.

The study of intermittency has been used to show multifractal aspects in the cascade dynamics behind multiparticle production. The dynamical cascade is connected to complex QCD diagrams which would describe the entire particle production process~\cite{Sarkisyan, Dremin2, greeks}. Here we show the connection between intermittency and Tsallis statistics. However a direct connection with the scattering dynamics governed by QCD is possible, as shown below.

\subsubsection{S-matrix and entropic index}

Another important result for thermofractals  is that the thermodynamical potential for the self-similar solution
\begin{equation}
 \Omega=\int_0^{\infty} \int_0^{\infty} A F^{\frac{3N}{2}} \exp\bigg(-\frac{\alpha F}{kT}\bigg) dF \bigg[1+(q-1)\frac{\varepsilon}{k\tau}\bigg]^{-1/(q-1)} d\varepsilon
\end{equation}
can be written in the form
 \begin{align}
 \Omega= & \Omega_o - \int_0^{\infty}  A \exp\bigg(-\frac{F}{kT}\bigg) F^{\frac{3N}{2}-1} \times \\
 & \bigg[1- \int_0^{\infty} \exp\bigg( -(q-1)\frac{\varepsilon}{Nk\tau}\frac{F}{kT} \bigg)\bigg(1+(q-1)\frac{\varepsilon}{k\tau}\bigg)^{-1/(q-1)} d\varepsilon \bigg] dF  \label{comparablepotential}
 \end{align}
where Eq.~(\ref{alfa}) was used and
 \begin{equation}
 \Omega_o= \int_0^{\infty} \int_0^{\infty} \exp\bigg(-\frac{F}{kT}\bigg) F^{3N/2} dF
 \end{equation}
 is the potential function for a non-interacting gas. Writing the potential in this form allows a direct comparison with the Dashen, Ma and Bernstein~\cite{DashenMaBernstein} formula connecting thermodynamics and microscopic information on the interaction among the particles composing the gas, which appears in terms of the scattering matrix, $S$, in
 \begin{equation}
 \Omega= \Omega_o - \frac{1}{4 \pi \beta i} \int_0^{\infty} \exp(-E/kT) \bigg(Tr S^{-1}\frac{\overset\leftrightarrow{\partial}}{\partial E}S \bigg)_C dE\,, \label{DashenMaBernstein}
\end{equation}
where the index $C$ indicates that the trace is performed for the connected diagrams in the Feynman-Dyson expansion.

Direct comparison of Eqs.~(\ref{comparablepotential}) and~(\ref{DashenMaBernstein}) gives
\begin{equation}
 \bigg(Tr S^{-1}\frac{\partial}{\partial E}S \bigg)_C \propto 1- F \int_0^{\infty} \exp\bigg(-\frac{(q-1)\varepsilon}{Nk\tau}\frac{F}{kT} \bigg)\bigg[1+(q-1)\frac{\varepsilon}{k\tau}\bigg]^{-\frac{1}{q-1}} d\varepsilon \label{Smatrix}
\end{equation}
which is a relation stablishing constraints in the $S$ matrix which will alow the interacting gas to show up nonextensive features. Equation~(\ref{Smatrix}) relates the $S$ matrix to the entropic factor, allowing one to extract information on the microscopic interaction from the non extensive behaviour of the experimental distributions.

\section{Conclusions}

The present work introduces a system which have a fractal structure in its thermodynamical functions, which is called thermofractal. It is shown that its thermodynamics is more naturally described by Tsallis statistics rather than the Boltzmann statistics. A relation between the fractal dimension and the entropic index, $q$, is found.
The ratio between the Tsallis temperature, $\tau$ and the Boltzmann temperature, $T$, is related to the entropic index and to the number of  subsystems, $N'$, in the next level of the fractal structure. It is shown that while $\tau$ regulates the system energy, $T$ regulates the fraction of the total energy that is accumulated in as internal energy of the subsystems. 

The study of the self-similar thermofractal reveals that it is a fractal with dimension determined by $q$ and $N'$. The Lipshitz-H\"older exponent is calculated in terms of $\tau$, $q$ and $N'$. Assuming that hadrons present a thermofractal structure, the relevant values for the calculation are obtained from the analyses of $p_T$ distribution and from the observed hadronic mass spectrum, while the ratio $\tau/T$ was already found in a work comparing the thermodynamical results to the LQCD data.

The comparison between the calculated fractal dimension and the value obtained from the analysis of intermittency in HEP experimental data show a fair agreement. This result is an indication that hadrons present a fractal structure similar to the thermofractal introduced here. Indeed, the calculated fractal dimension is obtained from a combination of $q$ and $\tau/T$ determined in analyses that are completely different of the analysis of intermittency.

Finally, for a system of interacting particles presenting thermofractal structure it is found a relation between the entropic index and the $S$-matrix for the particle interaction. This result allows on one hand to connect the entropic index to fundamental aspects of the interaction between the constituents, and on the other hand it establishes constraints on the $S$-matrix to allow the emergence of non extensivity in the corresponding system.

\section{Acknowlwdgements}
The author aknowedges fruitful discussions with D. Eugenio Meg\'ias (W. Heisenberg Institute - Max Plank Institute for Physics) and with Allbens Atman (Federal University of Minas Gerais). This work was supported by the Brazilian agency, CNPq, under grant 305639/2010-2 .


\begin{thebibliography}{1}
\bibitem{Tsallis1988} C. Tsallis, J. Stat. Phys. 52 (1988) 479.
\bibitem{Mandelbrot} B. B. Mandelbrot, The Fractal Geometry of Nature, W. H. Freeman, New York 1983.
\bibitem{Tel} T. T\'el, Z. Nat. 43a (1988) 1154.
\bibitem{Schroeder} M. Schroeder in Fractals, Chaos, Power Laws: Minutes from an Infinite Paradise, Dover Pub Inc. 1991 - Mineola NY.
\bibitem{LyraTsallis} L. Lyra and C. Tsallis, Phys. Rev. Lett. 80 (1998) 53.
\bibitem{Jizba} P. Jizba and T. Arimitsu, Ann. Phys. 312 (2004) 17.
\bibitem{Touchette} H. Touchette and C. Beck, J. Stat. Phys. 125 (2006) 459.
\bibitem{Olemskoi} A.I. Olemskoi, V.O. Kharchenko and V.N. Borisyuk, Physica A 387 (2008) 1895.
\bibitem{Fermi} E. Fermi, Prog. Theor. Phys. 5 (1950) 570.
\bibitem{Hagedorn65} R. Hagedorn, Nuovo Cimento Suppl. 3 (1965) 147.
\bibitem{Beck} C. Beck, Physica A: Statistical Mechanics and its Applications 286 (2000) 164 - 180.
\bibitem{WWselfsymmetry} G. Wilk and Z. W\l{}odarczyk, Phys. Lett. B (2013) 163.
\bibitem{Frautschi} S. Frautschi, Phys. Rev. D 3 (1971) 2821.
\bibitem{CabibboParisi} N. Cabibbo and G. Parisi, Phys. Lett. 59B (1975) 67.
\bibitem{Hagedorn83} R. Hagedorn, {\it Multiplicities, $p_T$-distributio and the expected hadron $\rightarrow$ quark-gluon phase transition}, Ref.TH.3684-CERN, 1983.
\bibitem{Bediaga2000156} I. Bediaga, E.M.F. Curado and J.M. de Miranda, Physica A: Statistical Mechanics and its Applications 286 (2000) 156 - 163.
\bibitem{Deppman} A. Deppman, Physica A 391 (2012) 6380.
\bibitem{Sena} I. Sena and A. Deppman, Eur. Phys. J. A (2013) 49: 17.
\bibitem{Cleyman_Worku} J. Cleymans and D. Worku, J. Phys. G: Nucl. Part. Phys. 39 (2012) 025006.
\bibitem{AzmiCleymans1} M. D. Azmi and J. Cleymans, J. Phys. G: Nucl. Part. 41 (2014) 065001.
\bibitem{AzmiCleymans2} M. D. Azmi and J. Cleymans, Eur. Phys. J. C 75 (2015) 430.
\bibitem{Lucas} L. Marques, E. Andrade-II and A. Deppman, Phys. Rev. D 87 (2013) 114022.
\bibitem{Alice2012} B. Abelev et al. (ALICE Collaboration), Eur. Phys. J. C 72 (2012) 2183.
\bibitem{CMS2011} V. Khachatryan et al. (CMS Collaboration), JHEP 05 (2011) 064.
\bibitem{WongWilk} C.-Y. Wong and G. Wilk, Phys. Rev. D 87 (2013) 114007.
\bibitem{Wilk2007} G. Wilk, Braz. J. Phys. 37 (2007) 714-716.
\bibitem{Bonasera} H. Zheng, L. Zhu and A. Bonasera, Phys. Rev. D 92 (2015) 074009.
\bibitem{Wong} C.-Y. Wong, G. Wilk, L. Cirto and C. Tsallis, Phys. Rev. D 91 (2015) 114027.
\bibitem{Borland} L. Borland, Phys. Lett. A 245 (1998) 67-72. 
\bibitem{Beck2} C.Beck, Physica A 331 (2004) 173.
\bibitem{Wilk2009} G. Wilk and Z. Wlodarczyk, Phys. Rev. C 79 (2009) 054903.
\bibitem{Biro} T. Bir\'o et al., Eur. Phys. J. A 49 (2013) 110.
\bibitem{BeckCohen} C. Beck, E. G. D. Cohen, Physica A 322 (2003) 267.
\bibitem{Thurner} R. Hanel and S. Thurner, Entropy 15 (2013) 5324.
\bibitem{Tempesta} P. Tempesta, Phys. Rev. E (2011) 021121.
\bibitem{Deppman2} A. Deppman, J. Phys. G: Nucl. Part. Phys 41 (2014) 055108.
\bibitem{Landau} L. Landau and E. Lifchitz in {\it Physique Statistique} Ed. MIR - Moscow 1967.
\bibitem{Megias} E. Meg\'ias, D. P. Menezes and A. Deppman, Physica A 421 (2015) 15.
\bibitem{Bialas_Peschanski} A. Bialas and R. Peschanki, Nucl. Phys. B273 (1986) 703.
\bibitem{Bialas_Peschanski2} A. Bialas and R. Peschanki, Nucl. Phys. B308 (1988) 857.
\bibitem{Hwa} R. C. Hwa, Phys. Rev. D 41 (1990) 1456.
\bibitem{HwaPan} R.C. Hwa and J. Pan, Phys. Rev. D 45 (1992) 1476.
\bibitem{DreminHwa} I,M. Dremin and R.C. Hwa, Phys. Rev. D 49 (1994) 5805.
\bibitem{Gang} X. Yi-Long, C. Gang, W. Juan, W. Mei-Juan and C. Huan, Nuclear Physics A 920 (2013) 33.
\bibitem{Gang2} C. Gang et al., Proc. XXXI Intern. Symp. Multiparticle Dynamics, Eds. B. Yuting, Yu Meiling, W. Yuantang, World Scientific, Singapore, 2002, p. 361.
\bibitem{Gang3} C. Gang, L. Lianshou and G. Yanmin, Int. J. Mod. Phys. A 14 (1999) 3687.
\bibitem{Ajienko} I.V. Ajienko et al., EHS/NA22 Collaboration, Phys. Lett. B 222 (1989) 306.
\bibitem{Rasool} M. H. Rasool and S. Ahamad, to apear in Chaos, Solitons and Fractals (2016).
\bibitem{SinghJain} G. Singh and P.L. Jain, Phys. Rev. C 50 (1994) 2508.
\bibitem{Albajar} C. Albajar et al., Z. Phys C 56 (1992) 37.
\bibitem{Ghosh} D. Ghosh et al., Phys. Rev. C 58 (1998) 3553.
\bibitem{Ahmad} A. Shafiq and M.A. Ahamad, Nucl. Phys. A 780 (2006) 206. 
\bibitem{LipaBuschbeck} P. Lipa and B. Buschbeck, Phys. Lett. B 223 (1989) 465.
\bibitem{Sarkisyan1} G. Alexander and E.K.G Sarkisyan, Nucl. Phys. B 92 (2001) 211.
\bibitem{Dremin2} I.M. Dremin, arxiv:hep-ph/9607346 (1996).
\bibitem{Sarkisyan} E.K.G. Sarkisyan, arxiv:hep-ph/0101218 (2001).
\bibitem{greeks} N.G. Antoniou, Y.F. Contoyiannis, F.K. Diakonos, Nucl. Phys. A 693 (2001) 799.
\bibitem{DashenMaBernstein} R. Dashen, S. Ma and H. J. Bernstein, Phys. Rev. 187 (1969) 345.

  
\end{thebibliography}
\end{document}